\documentclass[twocolumn,showpacs,preprintnumbers,amsmath,amssymb]{revtex4}
\usepackage{graphicx}
\usepackage{amssymb}
\usepackage[latin1]{inputenc}
\usepackage{dcolumn}
\usepackage{bm}

\begin{document}


\title{Electronic States of  Wires and Slabs of Topological Insulators: Quantum Hall Effects and Edge Transport. }

\author{L. Brey$^{1}$ and H.A.Fertig$^{2}$}
\affiliation{$^1$ Instituto de Ciencia de Materiales de Madrid, (CSIC),
Cantoblanco, 28049 Madrid, Spain \\
$^2$Department of Physics, Indiana University, Bloomington, IN 47405}

\date{\today}


\begin{abstract}

We develop a simple model of surface states for topological insulators, developing matching
relations for states on surfaces of different orientations.  The model allows one to write
simple Dirac Hamiltonians for each surface, and to determine how perturbations that couple
to electron spin impact them.  We then study two specific realizations of such systems:
quantum wires of rectangular cross-section and a rectangular slab in a magnetic field.
In the former case we find a gap at zero energy due to the finite size of the system.
This can be removed by application of exchange fields on the top and bottom surfaces,
which lead to gapless chiral states appearing on the lateral surfaces.  In the presence
of a magnetic field, we examine how Landau level states on surfaces perpendicular to the
field join onto confined states of the lateral surfaces.  We show that an imbalance in
the number of states propagating in each direction on the lateral surface is sufficient
to stabilize a quantized Hall effect if there are processes that equilibrate the
distribution of current among these channels.

\end{abstract}
\maketitle

\section{Introduction.}
Topological insulators (TI) are materials with insulating bulks and conducting
surfaces. These materials typically are gapped band insulators,
where strong spin-orbit coupling has inverted the usual energetic
ordering of the bands.  Near an interface
with the vacuum the bands revert to their usual order, inducing
two-dimensional metallic states on their surfaces \cite
{hasan_2010,Qi_2011,YAndo_2013}.
These surface states have a conical dispersion, and are described by a two dimensional massless Dirac equation centered at a time reversal
invariant point in momentum space. The metallic character of these states
is protected by time-reversal symmetry, so that a gap can only be
opened by perturbations which break this symmetry, e.g. magnetic
or exchange fields.
Angle-resolved photoemission spectroscopy has confirmed the existence of these surface states in certain materials \cite{Hsieh_2008}.


A magnetic field $B$ applied perpendicular to the surface of the topological insulator quantizes the orbital motion of the electrons and reorganizes
the energy spectrum into Landau levels with energies
\cite{McClure_1956,Cheng_2010,Hanaguri_2010,Jiang_2012,Taskin_2012}
\begin{equation}
E_{\pm} = \pm \sqrt{2n}  \hbar \omega_c,
\label{LL_Dirac}
\end{equation}
where $ \omega _c =  v _F /\ell $, $v_F$ is the speed of the carriers
at the Dirac points in the absence of the field,
$\ell=\sqrt {\frac {\hbar c}{e B}}$ is the magnetic length, and $n=0,1,2,...$.
If the Fermi energy of the system passes through the Dirac point for $B=0$,
the corresponding Fermi energy for $B \ne 0$
is pinned in the $n=0$ Landau level, such that half the states in that
level are occupied \cite{Jackiw_1984}.  Particle-hole symmetry
suggests that the Hall conductivity vanishes in this situation.
If the Fermi level is raised from this zero point, as it passes between
the $n-{\it th}$ and the $(n+1)$-$th$ Landau levels the Hall conductivity
from a single surface Dirac point is then quantized to
\begin{equation}
\sigma _{xy} =(n+\frac 1 2) \frac {e^2} h \, \, .
\label{half-integer}
\end{equation}

In general,
the integer quantized Hall effect is associated with current-carrying chiral edge modes. Each mode contributes $\pm e^2/h$ to the Hall conductivity when the Fermi
energy passes through it, so one expects only $integer$ quantization is possible,
in contrast to the bulk result suggested by Eq. \ref{half-integer}.
The resolution of this discrepancy  is that in actual samples, whatever the geometry may be, different surfaces are always connected, and may share
edge modes \cite{Fu_2007,Lee_2009,Vafek_2011,YYZhang_2012}.
Contacts used to measure the Hall conductance
will thus inevitably probe Dirac cones on more than one surface, so
that the observed $\sigma_{xy}$ is always quantized in integer units
of $e^2/h$.



A Hall effect may occur in a system whenever
there is time reversal symmetry-breaking, so that it may be induced in
principle without an external applied magnetic field.
Doping the system with magnetic impurities, or placing a surface
in proximity to a ferromagnetic insulator which can exchange-couple to
the TI surface, offer two non-standard methods to induce $\sigma_{xy} \ne 0$.
In such systems, the effective surface Hamiltonian in the absence of such
perturbations has the form $H = \hbar v_F( \sigma _x k_y- \sigma _y k_x)$,
in which the Pauli matrices $\sigma_x$, $\sigma_y$ represent electron spin operators.
In the presence of an exchange field pointing in the $\hat{z}$ direction
a gap opens in the surface spectrum, and a Chern number associated with
the states on such a surface has the value $\pm 1/2 e^2 / h$, with sign determined by the direction of the magnetization \cite{Yokoyama_2010,Garate_2010}.
This leads to an {\it  anomalous half integer} contribution of the
surface to the Hall conductivity,
$\sigma _{xy} = \pm e ^2/2h$  \cite{Qi_2008,Qi_2009,Yu_2010}.
As in the usual quantum Hall effect,
the existence of multiple surfaces in any real geometry
prevents a direct observation of a half-integral quantized Hall
conductance in such systems: measured Hall conductances are
always integrally quantized. An effect much like this has recently been observed in
thin films of (Bi,Sb)$_2$Te$_2$
doped with Cr atoms \cite{Chang_2013}.

In this work we introduce a simplified approach to analyzing transport in
TI surfaces which are nominally flat, but where surfaces of different
orientations may be connected.  We show that these simplified surface
states give a good accounting of their dispersion within the bulk gap,
and develop matching conditions for surfaces of different orientations.
We then apply the formalism first to the problem of a quantum wire with
rectangular cross-section, examining the effect of different exchange fields
on opposite surfaces.  In particular, we find for equal exchange fields that
the quantum wire supports  gapless chiral states, but when oriented oppositely
these states vanish and there is a gap in the spectrum.

In a second application, we consider the quantum Hall problem for a rectangular
slab geometry, with field oriented perpendicular to one pair of surfaces.
We show that the bulk Landau levels couple surface states on sides parallel
to the magnetic field, arriving at results very similar to those of
Ref. \onlinecite{Vafek_2011}.  We then analyze the effect of phase-breaking
processes by contacting the system to equilibrating voltage probes,
and argue that in this circumstance the system should support a quantized Hall effect.


This paper is organized as follows. In Section II we introduce the three dimensional Hamiltonian that describes the low energy properties of a  prototypical TI such as Bi$_2$Se$_3$. In Section III we obtain Dirac-like Hamiltonians that describe the various surface states.  Section IV is
devoted to a discussion of the matching conditions joining states on
different surfaces.  In Section V we discuss the energy spectrum of
the rectangular quantum wire, and Section VI discusses what happens to
this when exchange fields are introduced on two parallel surfaces.
In Section VII we consider the case of TI surfaces in a magnetic field,
discuss the energy spectrum in the quantum Hall regime, and then
analyze transport in this system in a multi-terminal geometry
with phase-breaking leads.  We conclude with a summary in Section VIII.

\section{Bulk Hamiltonian}

\begin{figure}
 \includegraphics[clip,width=8.5cm]{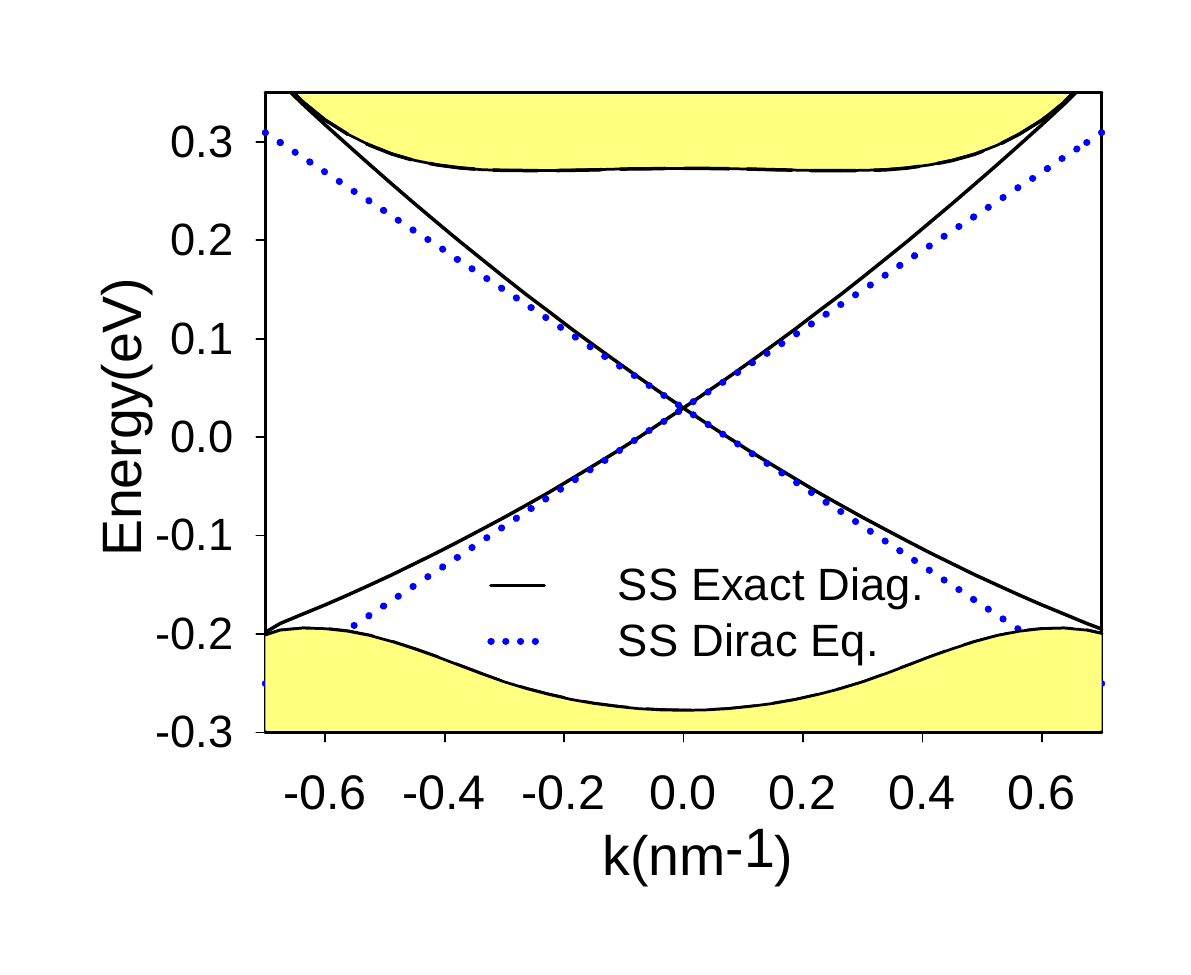}
 \caption{(Color online) Band structure of a thick TI slab, normal  to the $\hat z$ direction, of obtained from diagonalizing the Hamiltonian in Eq. \ref{H3D}. Shadow region represents the bulk band structure.  The states in the gap correspond to surface states. Dotted lines represent the surface states obtained from the Dirac Hamiltonian Eq. \ref{Hz}.}
 \label{Fig1}
\end{figure}

The  properties of three dimensional topological
insulators in the Bi$_2$Se$_3$ family of materials can be
described by a four band Hamiltonian introduced by Zhang {\it et
al.} \cite{Zhang_2009}. In the ${\bf k }\cdot {\bf p }$
approximation, states near zero energy
are controlled by an effective continuum Hamiltonian of the form
\begin{equation}
H^{3D}  =  E({\bf k}) +
\left(
  \begin{array}{cccc}
     \mathcal{M} ( {\bf k} ) & A_1 k_z & 0 & A_2 k _ - \\
    A _1 k_z   & -\mathcal{M}( {\bf k} ) & A_2 k _ - & 0 \\
    0 & A_2 k _ +  &   \mathcal{M} ( {\bf k} ) & - A _1 k _z \\
    A_2 k _ +  & 0 & -A _1 k _z  & - \mathcal{M}( {\bf k} ) \\
  \end{array}
\right),
\label{H3D}
\end{equation}
where $\mathcal{M} ( {\bf k} )$=$M_0 - B_2 (k_x ^2 + k_y ^2)-B_1k_z
^2$, $k_{\pm}$=$k_x \pm i k_y$ and  $E({\bf k})$=$ C + D _1 k_z ^2 +
D_2 (k_x ^2+k_y ^2)$. The basis states for which this Hamiltonian is written are
$|1>$=$|p1 _z ^+, \uparrow>$, $|2>$=$-i |p2 _z ^- ,
\uparrow>$,$|3>$=$|p1 _z ^+, \downarrow>$, and $|4>$= $i|p2 _z ^-,
\downarrow>$, which are hybridized states of the Se $p$ orbitals
and the Bi $p$ orbitals, with even $(+)$ and odd $(-)$ parities, and
spin up ($\uparrow$) and down ($\downarrow$). The Hamiltonian
parameters for a particular material can be obtained by fitting to {\it ab initio} band structure calculations \cite{Liu_2010}. In the case of Bi$_2$Se$_3$
the relevant parameters are
$M_0$=0.28$eV$, $A_1$=2.2$eV \AA $, $A_2$=4.1$eV\AA$, $B_1$=10$eV \AA ^2$, $B_2$=56.6$eV\AA^2$,
$C$=-0.0068$eV$, $D_1$=1.3$eV \AA ^2$ and $D_2$=19.6$eV\AA^2$.
In Fig. \ref{Fig1} we plot the band structure of a thick slab of 
topological insulator with these parameters.
The electronic structure is obtained by diagonalizing   
Eq. \ref{H3D} with $k_z \rightarrow -i \partial _z$, for fixed $k_x$ and $k_y$, using basis
states in which the wavefunctions to vanish at the surfaces of the slab \cite{Lasia_2013}.
The system is rotationally invariant in the $x$-$y$ plane, so that 
in Fig.\ref{Fig1} $k$ represents the magnitude of the in-plane momentum.


In what follows, we will be interested in coupling the spin degree of freedom
to effective magnetic fields, created by exchange coupling to magnetic
insulators or magnetized impurities.  To do this we need the spin
operators in the basis of bulk states.  These are \cite{Silvestrov_2012}
\begin{eqnarray}
  \! \! S_x  = \left(
       \begin{array}{rcrc}
        \! 0 & \!0 & \!1 & \!0 \\
        \! 0 & \!0 & \!0 & \!-1 \\
        \!1 & \!0 & \!0 & \!0 \\
        \! 0 & \!-1 & \!0 & \!0 \\
       \end{array}
     \right) \!   , \nonumber \\
\!\! S_y = \left(
       \begin{array}{rccr}
        \!0 & \!0 & \!-i & \!0 \\
        \! 0 & \!0 & \!0 & \!i \\
        \!i & \!0 & \!0 & \!0 \\
         \!0 &\!-i & \!0 & \!0 \\
       \end{array}
     \right), \nonumber
\end{eqnarray}
and
\begin{eqnarray}
\!\! S_z = \left(
       \begin{array}{rrcc}
        \! 1 & \!0 & \!0 &\! 0 \\
        \! 0 & \!1 & \!0 & \!0 \\
        \! 0 & \!0 & \!-1 & \!0 \\
        \! 0 & \!0 & \!0 & \!-1 \\
       \end{array}
     \right). \nonumber \\
     \label{3Dspin}
\end{eqnarray}
We will project these operators onto surface states,
yielding operators which depend on the orientation of the surface
with respect to the bulk {\bf k} axes.  It is important to take into
account their precise form when analyzing the influence of an
effective Zeeman field at such surfaces.

\section{Surface Hamiltonians}
An important feature of $H^{3D}$ is that, due to its non-trivial
topology, when a surface is introduced
one finds states in the gap which can be represented by Dirac Hamiltonians.
In this section we will write down explicit forms for these surface
states, following an approach introduced by Silvestrov and coworkers
\cite{Silvestrov_2012}, albeit in a simplified form which allows
an introduction of simple matching conditions between different surfaces.
Because of
the strong anisotropy of these layered materials, for a given surface the states depend on the orientation of that surface.  We confine our analysis to
surfaces of high symmetry ($\hat{x}, \hat{y}, \hat{z}$),
and define the surface orientation by normal vectors
$\hat n = \pm\hat x,\,\hat n = \pm\hat y$ and $\hat n = \pm\hat z$.
Generally speaking, the strategy is to find states which vanish
in all its components on the surface, are evanescent
as one moves into the bulk of the system, and are constant
on planes of constant depth into the bulk.  Such states have energies
within the bulk gap, and generically one finds two such states
which are degenerate.  Following the \hbox{{\bf k} $\cdot$ {\bf p}} approximation,
one assumes a good approximation to states near this energy can be
formed out of linear combinations of these bound surface states
(envelope functions)
multiplying plane wave states with wavevector parallel to the surface,
and projects the bulk Hamiltonian, Eq. \ref{H3D}, into the space
of these two states.  This results in a $2\times 2$ Hamiltonian,
with a Dirac spectrum
in the absence of other perturbations.
Appendix A details, as an example, how one
obtains the surface Hamiltonian for $\hat{n}_z$.  In the following
subsections we present the results of such calculations for the
three orientations.

\subsection{$\pm  \hat z$ Surface}
The Hamiltonian describing the electrons moving on the surface
with $\hat n = \pm \hat z$ has the form
\begin{equation}
H^{ \pm \hat z} \!  = \! \frac {D_1}{B_1}M_0
 \! \pm \!  A_2 \sqrt{1\! - \! \frac {D_1^2}{B_1^2}} \left(
  \begin{array}{cc}
     0  &  i k_x \! +\! k_y \\
      -i k_x \! + \!k_y  &0 \\
 \end{array}
\right) .
\label{Hz}
\end{equation}
This Hamiltonian describes two dimensional Dirac fermions with velocity
$v_F=A_2 \sqrt{1\! - \! \frac {D_1^2}{B_1^2}}$. In  Fig.\ref{Fig1} we compare the dispersion obtained from the Dirac Hamiltonian Eq.\ref{Hz}, $ \varepsilon = \frac {D_1}{B_1}M_0 \pm
v_F \sqrt{k_x^2+k_y ^2} $,  with the exact result obtained from the diagonalization of the 3D Hamiltonian in a thick slab geometry.  In the region near the Dirac point, where the dispersion is linear, the Dirac Hamiltonian
yields a good description of the surface band structure.

The two states resulting from the solution of the surface
problem, which are the envelope functions used in the basis
of Eq.\ref{Hz}, are
\begin{equation}
u  ^{\pm \hat z}\! \! = \! \!\frac 1 {\sqrt{2}} \left ( \begin{array} {c}     \sqrt{ 1 \!  + \! \frac {D_1}{B_1}} \\   \! \! \! \mp i\sqrt{ 1\! -\! \frac {D_1}{B_1}}\\0 \\ 0 \end{array}\right )
 \, , \,
v ^{\pm \hat z} \! \!=\!  \!\frac 1 {\sqrt{2}} \left ( \begin{array} {c}     0\\ 0 \\  \sqrt{ 1\! + \! \frac {D_1}{B_1}}  \\ \! \! \! \pm  i\sqrt{ 1 \!- \!\frac {D_1}{B_1}} \end{array}\right ).
\end{equation}

For this surface orientation
the electron spin operators, formed by projecting the
the full spin operators (Eq. \ref{3Dspin}) onto the two
surface states,
coincide with the Pauli spin matrices,
\begin{equation}
S_x=\sigma_x \, \, , \, \, S_y=\sigma _y \, \, {\rm and} \, \, S_z = \sigma _z  .
\label{spinz}
\end{equation}
Thus, magnetic impurities  or the proximity of ferromagnetic insulators will open a gap in the Dirac spectrum only when their magnetization
has a component in the $\hat z$ direction.  As mentioned in the Introduction,
Eq. \ref{Hz} in this case picks up a Dirac mass term.  The integral of
the Berry's curvature in the vicinity of the (now gapped) Dirac point
then becomes half integral, and the resulting contribution to the Hall
conductivity of electrons in these states is half-integral
\cite{Yokoyama_2010,Garate_2010}.

\subsection{$\pm  \hat  x$ Surface}
The Hamiltonian describing the electrons moving in the surface perpendicular
to the $\pm \hat y$ direction has the form
\begin{equation}
H^{ \pm \hat x}  = \frac {D_2}{B_2}M_0  \mp  \sqrt{1-\frac {D_2^2}{B_2^2}} \left(
  \begin{array}{cc}
     A_2 k_y  &  iA_1 k_z \\
      -i A_1 k_z &-A_2 k_y\\
 \end{array}
\right) .
\label{Hx}
\end{equation}
The envelope states forming the basis of this Hamiltonian are
\begin{equation}
u  ^{\pm \hat x}\! \! = \! \!\frac 1 {\sqrt{2}} \left ( \begin{array} {c}     \sqrt{ 1 \!  + \! \frac {D_2}{B_2}} \\   0 \\ 0 \\ \! \! \! \mp i\sqrt{ 1\! -\! \frac {D_2}{B_2}} \end{array}\right )
 \, , \,
v ^{\pm\hat x} \! \!=\!  \!\frac 1 {\sqrt{2}} \left ( \begin{array} {c}     0\\ \mp  i   \sqrt{ 1\! - \! \frac {D_2}{B_2}}  \\ \! \! \!  \sqrt{ 1 \!+ \!\frac {D_2}{B_2}} \\0 \end{array}\right ).
\label{basisx}
\end{equation}
For this surface orientation the projection of the spin operators in Eq. \ref{3Dspin} become
\begin{equation}
S_x=\frac {D_2} {B_2}\sigma_x  \, \, , \, \, S_y=\sigma _y \, \, , \, \, S_z = \frac {D_2} {B_2} \sigma _z .
\label{spinx}
\end{equation}
The non-integral coefficients in $S_x$ and $S_z$ arise because the
envelope states, Eq. \ref{basisx},  have non-zero amplitudes for
microscopic orbitals
with different spin orientations.
As in the case of the $\pm {\hat z}$ surface, magnetic impurities polarized
with a component in the normal direction to the surface open a gap in the Dirac
spectrum. Note that for this surface,  $S_x \propto \sigma _x$ because of the diagonal term  $E({\bf k})$ in Eq. \ref{H3D}; for $D_2=0$, on this surface
the component of spin in the $x$ direction will always be zero.

\subsection{$\pm  \hat  y$ Surface}
The projected states and resulting Hamiltonian
for the $\pm \hat y$ surfaces are qualitatively very
similar to those of the $\pm \hat{x}$ surfaces.
The Hamiltonian has the form
\begin{equation}
H^{ \pm \hat y}  = \frac {D_2}{B_2}M_0
 \pm  \sqrt{1-\frac {D_2^2}{B_2^2}} \left(
  \begin{array}{cc}
     A_2 k_x  &  -A_1 k_z \\
       -A_1 k_z &-A_2 k_x\\
 \end{array}
\right) .
\label{Hy}
\end{equation}
The envelope states for the states in which the Hamiltonian, Eq. \ref{Hy}, is
expressed are
\begin{equation}
u^{\pm \hat y}\! \! = \! \!\frac 1 {\sqrt{2}} \left ( \begin{array} {c}     \sqrt{ 1 \!  + \! \frac {D_2}{B_2}}   \\ 0 \\ 0 \\ \! \! \! \pm \sqrt{ 1\! -\! \frac {D_2}{B_2}}   \end{array}\right )
 \, , \,
v ^{\pm \hat y} \! \!=\!  \!\frac 1 {\sqrt{2}} \left ( \begin{array} {c} 0 \\      \mp     \sqrt{ 1\! - \! \frac {D_2}{B_2}}  \\ \! \! \!  \sqrt{ 1 \!+\!\frac {D_2}{B_2}} \\0  \end{array}\right ).
\label{basisy}
\end{equation}
Finally, the projections of the spin operators, Eq. \ref{3Dspin},
are in this case
\begin{equation}
S_x=\sigma_x  \, \, , \, \, S_y=\frac { D_2}{B_2} \sigma _y \, \, {\rm and} \, \, S_z = \frac {D_2} {B_2} \sigma _z .
\end{equation}

\section{Matching Conditions}
As discussed in the Introduction, in many situations one cannot understand the transport
properties of a TI based on individual surfaces in isolation; it is necessary
to understand how these surfaces connect.  Towards this end, in this
section we develop a
simple approach to matching wavefunctions on a line junction separating
two perpendicular surfaces, labeled $1$ and $2$, of a three-dimensional TI.  We assume
the Fermi energy is in a bulk gap and focus on how these matching conditions
impact the surface state spectra and associated conduction properties.
Our method uses a general approach \cite{Londergan_Book},
in which wavefunctions of a system are matched along some chosen surface
that divides the system into disparate pieces, each having different
transverse modes into which it is natural to decompose wavefunctions.
In principle the matching can be carried out precisely by considering overlaps
along the chosen surface
of all the transverse modes.  In practice it is
usually necessary to truncate the number of modes kept, allowing one to obtain
approximate results for wavefunctions in some energy interval.  This
general approach has been quite successful for treating semiconductor nanostructures
\cite{Londergan_Book}, and recently has been useful for understanding
transport through graphene nanostructures \cite{Iyengar_2008}.

In the present
context we are interested in understanding spectra and transport when
there are only a small number of modes crossing the Fermi energy,
with wavefunctions confined to the surfaces, while all other transverse
modes (associated with the bulk) represent states well above or below
the Fermi energy.  These latter states are incorporated as
evanescent states which do not directly contribute to the current in
the system, although they quantitatively affect the scattering
among conducting modes.  The simplest approximation in this situation
is to ignore the evanescent modes entirely, leading to an
``open mode approximation'' \cite{Londergan_Book}.  However, this does
not define the approximation scheme uniquely, as one may
choose the matching surface to obtain the best results.  Below we
demonstrate that demanding that the truncated Hamiltonian be
Hermitian effectively singles out a specific set of matching conditions within
the open mode approximation.

\subsection{Open Mode Approximation}

\begin{figure}
 \includegraphics[clip,width=7.5cm]{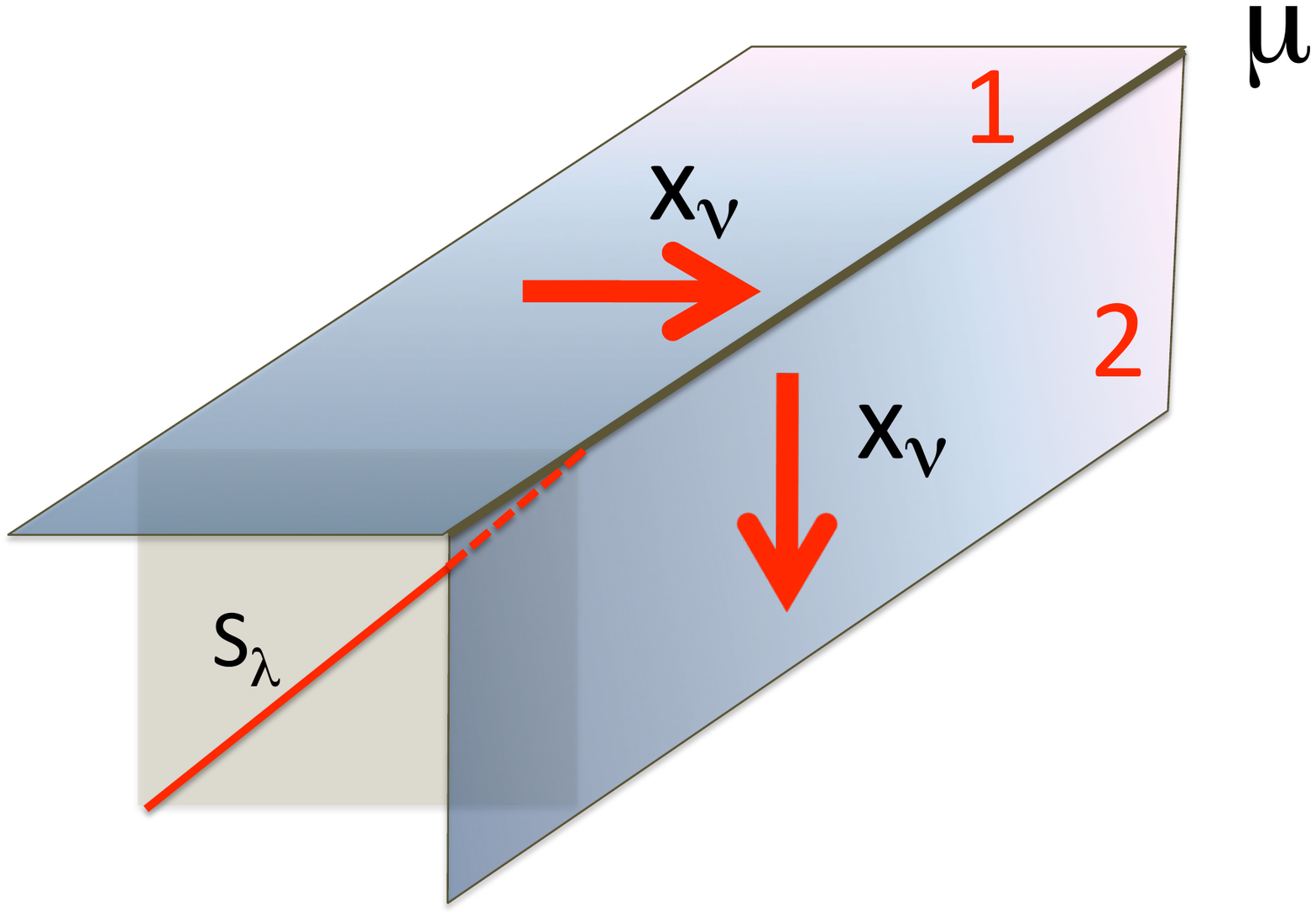}
 \caption{(Color online) Junction between two perpendicular surfaces, $1$ and $2$ of a three-dimensional TI.
 The direction of the line junction is $\mu$.  $x_{\nu}$ indicates the  effective one dimensional coordinate  in the surfaces $1$ and $2$. $s_{\lambda}$ parameterizes a 
 curve perpendicular to $\mu$ (here a straight line) on the surface joining the two 
 slabs at a corner junction.}
 \label{Fig2}
\end{figure}

To motivate our matching conditions, it is useful to consider two slabs
of the TI system, with surface normals perpendicular to one another,
joined through a perpendicular junction.  Fig. \ref{Fig2} illustrates
the corner of such a junction, emphasizing the role of the
surface states, which are most important when the Fermi energy is
in the bulk gap.  For concreteness we assume one of these has
horizontal surfaces, perpendicular to $\hat z$, and the other
vertical surfaces perpendicular to $\hat x$.  Assuming that the
system is uniform along the $\hat y$ direction so that $k_y$
is a good quantum number,
states of the first slab
can be written in the form
$\Psi^{(z)}=e^{ik_y y} \sum_n e^{ik_x^{(n)} x} c_n^{(z)} \chi^{(z)}_n(z)$,
and for the second slab
$\Psi^{(x)}=e^{ik_y y} \sum_n e^{ik_z^{(n)} z} c_n^{(x)} \chi^{(x)}_n(x)$.
In these expressions
$\chi_n^{(z,x)}$ are transverse wavefunctions for the
slabs, among which are the surface states discussed in the last
section \cite{comment_surface_states} and for this problem are
four component vectors; the wavevectors $k_{x,z}^{(n)}$ are
determined by the energy of the state, and  in most cases are actually
complex (i.e., represent evanescent states)
if the Fermi energy is in the bulk band gap, and the coefficients $c_n^{(x,z)}$
are weights which must be related by appropriate matching conditions.
This last requirement in principle can be implemented by matching all components of
the wavefunctions on some surface along which the two
slabs are joined together.  In principle one may choose any convenient
surface, and parameterize it as $(x_{\lambda},z_{\lambda})$, with
$0 \le \lambda \le 1$.  For a given set of coefficients on one
side of the junction, say $\{ c_n^{(z)} \}$, the coefficients on the
other side can in principle be found \cite{Londergan_Book}
by a matrix multiplication,
$c_n^{(x)} = \sum _m <x,n|z,m> c_m^{(z)}$ with
\begin{eqnarray}
<x,n|z,m> = \,\,\,\,\,\,\,\,\,\,\,\,\,\,\,\,\,\,\,\,\,\,\,\,\,\,\,\,\,
\,\,\,\,\,\,\,\,\,\,\,\,\,\,\,\,\,\,\,\,\,\,\,\,\,\,\,\,\,
\,\,\,\,\,\,\,\,\,\,\,\,\,\,\,\,\,\,\,\,\,\,\,\,\,\,\,\,\,
\nonumber \\
\int_{\lambda=0}^{\lambda=1} ds_{\lambda}
e^{-ik_z^{(n)}z_{\lambda}+ik_x^{(m)}x_{\lambda}}
<\chi_n^{(x)}(x_{\lambda})|\chi_m^{(z)}(z_{\lambda})>. \nonumber
\end{eqnarray}
Here $ds_{\lambda}$ is the differential arc length along the joining surface,
and $<\chi_n^{(x)}(x)|\chi_m^{(z)}(z)>$ is the dot product of the
four component vectors.

In general, the challenge in carrying out this matching is that
the full matrix $<x,n|z,m>$ is difficult to compute.  Moreover,
typically most of the transverse modes are
``closed'' -- i.e., they host evanescent states --
and do not contribute directly to current across the junction.
In the ``open mode'' approximation one simply ignores the closed modes
and retains only those that are current-carrying at the Fermi energy.
In the present context this is particularly simplifying since only
the surface modes are open when the Fermi energy is in the bulk gap.

In the present case it is then natural to retain only the $u$ and $v$ modes detailed
in the last section.  If one further assumes that the penetration depths
of the surface states ($\lambda_1$ and $\lambda_2$ in the Appendix)
are short, such that the phase factor $e^{-ik_z^{(n)}z_{\lambda}+ik_x^{(m)}x_{\lambda}}$
has a negligible variation on the joining surface in the region
where
$<\chi_n^{(x)}(x_{\lambda})|\chi_m^{(z)}(z_{\lambda})>$
is significantly different than zero, the resulting connection
between coefficients takes the simple form
\begin{equation}
\left ( \begin{array} {c}     c^{x}_u \\  c^{x}_v   \end{array} \right ) = {\Large M} _{x,z}
\left ( \begin{array} {c}     c^{z}_u \\  c^{z}_v  \end{array}\right ) \, \, ,
\label{Mxz}
\end{equation}
where we have written the two open channel coefficients for each surface
$c^{\mu}_n$ with $n \rightarrow u,v$.  The matrix ${\Large M} _{x,z}$
has the form
\begin{eqnarray}
& {\Large  M } _{x,z}    =    {\cal C}_{xz} \!
\left ( \begin{array} {cc} \gamma _+ &  \gamma _-\\  -\gamma _-  &  \gamma _+
  \end{array}\right )
\label{Mxzform}
\end{eqnarray}
with
\begin{equation}
\gamma_{\pm}  = \sqrt{ [ (1\pm  D_{1}/B_{1})( 1 \pm  D_{2}/B_{2}) ] },
\nonumber
\end{equation}
and
$
{\cal C}_{xz}=\int ds_{\lambda} [e^{\lambda_1^xx_{\lambda}}-e^{\lambda_2^xx_{\lambda}}]
[e^{\lambda_1^zz_{\lambda}}-e^{\lambda_2^zz_{\lambda}}],
$
with $\lambda_{(1,2)}^{(x,z)}$ the corresponding $\lambda_{(1,2)}$
constants in the Appendix.

At this level of approximation, the only relevant information about the joining
surface is contained in the constant ${\cal C}_{xz}$. We thus will ultimately
choose this constant -- implicitly, by choosing the joining surface -- to
obtain the best approximation, which we will argue below leaves the projected
Hamiltonian Hermitian; this choice uniquely fixes the value of the constant.
Before turning to this, we summarize the results of the open mode approximation
for other possible 90$^\circ$ corner junctions with surfaces normal to principal
axes of the structure.  In general, we write
\begin{equation}
\left ( \begin{array} {c}     c^{\mu}_u \\  c^{\mu}_v   \end{array} \right ) = {\Large M} _{\mu,\nu}
\left ( \begin{array} {c}     c^{\nu}_u \\  c^{\nu}_v  \end{array}\right ) \, \, ,
\label{Mat_Con}
\end{equation}
with $\mu,\nu=x,y,z$, and
\begin{eqnarray}
& {\Large  M } _{z,x}    =  {\Large M} _{-z,-x} =   {\cal C}_{xz} \!
\left ( \begin{array} {cc} \gamma _+ &  \gamma _-\\  -\gamma _-  &  \gamma _+
  \end{array}\right )
\nonumber \\
& {\Large M}  _{-z,x}   =   {\Large M} _{z,-x} = { \cal C}_{zx} \!
\left ( \begin{array} {cc} \gamma _+ &  -\gamma _-\\  \gamma _-  &  \gamma _+
  \end{array}\right )
\nonumber \\
& {\Large M}  _{z,y}    =   {\Large M} _{-z,-y} =   {\cal C}_{yz} \!
\left ( \begin{array} {cc} \gamma _+ & i  \gamma _-\\  i\gamma _-  &  \gamma _+
  \end{array}\right )
\nonumber \\
& {\Large M}   _{z,-y}    =  {\Large M} _{-z,y} =   {\cal C}_{zy} \!
\left ( \begin{array} {cc} \gamma _+ & - i \gamma _-\\ - i\gamma _-  &  \gamma _+
  \end{array}\right )
\nonumber \\
& {\Large M}   _{\pm x,\pm y }     = {\cal C}_{xy}
\left ( \begin{array} {cc} 1 & 0\\ 0  &  1
  \end{array}\right ).
  \end{eqnarray}

\subsection{Hermitian Effective Hamiltonian}
As discussed above, we would like to choose the $\cal{C}_{\mu,\nu}$ coefficients
to optimize the approximation.  In particular, in order to obtain sensible
results within the approximation scheme, the projected Hamitonian of the
full system should be Hermitian.  This guarantees among other things that
current will be conserved across the junctions.  We now show that this
requirement uniquely fixes the coefficients $\cal{C}_{\mu,\nu}$.

As a concrete example we return to the geometry illustrated in
Fig. \ref{Fig2}. The system is invariant along the $\hat{y}$ direction,
so that we can consider the system for each $k_y$ as one-dimensional,
with a single coordinate along the surface, running perpendicular to the
line junction.  The corner can be ``flattened'' by taking $x<0$ to
represent the $\hat{z}$ surface, and $x>0$ to represent the $\hat{x}$ surface,
which we refer to respectively as the 1 and 2 surfaces in what follows.
In this notation, the portion of
the low energy Hamiltonian which represents the problem has the form
\begin{equation}
h_x= i\hbar A(x)\sigma_y \partial_{x}\, .
\label{hx}
\end{equation}
$A(x)$ is piecewise constant but jumps at $x=0$.
Potentially this leads to problems because matrix elements
between arbitrary two-component wavefunctions $\phi_1(x)$
and $\phi_2(x)$ may not
obey $\int dx \phi_1^* h_x \phi_2 = \int dx (h_x\phi_1)^* \phi_2$
due to a surface term at $x=0$ from integration by parts.
In particular \cite{Vafek_2011}, the Hamiltonian is only Hermitian if
\begin{equation}
A^+\phi_1^{-\dag}\sigma_y\phi_2^-=A^-\phi_1^{+\dag} \sigma_y \phi_2^+,
\label{hermitian}
\end{equation}
where $A^{\pm} \equiv A(x=0^\pm)$ and $\phi_{1,2}^{\pm} \equiv \phi_{1,2}(0^{\pm})$.
{}From Eq. \ref{Mxz}, this means
$$
A^+\phi_1^{-\dag}\sigma_y\phi_2^-=A^-\phi_1^{+\dag} \sigma_y
{\Large M}_{xz} \phi_2^-,
$$
from which we read off
$$
A^+\phi_1^{-\dag}\sigma_y=A^-\phi_1^{+\dag} \sigma_y {\Large M}_{xz}.
$$
Taking the Hermitian conjugate of this yields
$$
A^+\sigma_y \phi_1^{-} = A^- {\Large M}_{xz}^{\dag} \sigma_y \phi_1^{+},
$$
and since $\phi_1^{+}={\Large M}_{xz}\phi_1^{-}$, we arrive at
the relation
\begin{equation}
{{A^+}\over{A^-}} = \sigma_y {\Large M}_{xz}^{\dag} \sigma_y {\Large M}_{xz}.
\label{matrix_hermitian}
\end{equation}
{}From the form of Eq. \ref{Mxzform} we see $M_{x,z}={\cal C}_{xz} (\gamma_+
+i \gamma_- \sigma_y)$, and plugging this into Eq. \ref{matrix_hermitian}
above, we arrive at the condition
\begin{equation}
{\cal C}_{xz}^2= {{A^+}\over{A^-}}(\gamma_+^2+\gamma_-^2)^{-1}.
\label{Cxz}
\end{equation}
For the $xz$ line junction, $A^+=A_1$ and $A^-=A_2$.

Eq. \ref{Cxz} uniquely specifies the matching condition we should use
in an open mode approximation to get physically sensible results.
It is interesting to note that if one sets $\phi_1=\phi_2$ in Eq. \ref{hermitian},
the resulting condition is precisely what is needed to get current
conservation across the junction.  Finally, generalizing this result
to other corner junctions, we find
\begin{eqnarray}
{\cal C}_{zx}^2 &=& {{A_2}\over{A_1}}(\gamma_+^2+\gamma_-^2)^{-1} \nonumber \\
{\cal C}_{yx}^2 &=& {{A_1}\over{A_2}}(\gamma_+^2+\gamma_-^2)^{-1} \nonumber \\
{\cal C}_{zy}^2 &=& {{A_2}\over{A_1}}(\gamma_+^2+\gamma_-^2)^{-1} \nonumber \\
{\cal C}_{xy} &=& 1.
\label{coefficients}
\end{eqnarray}


\begin{figure}
 \includegraphics[clip,width=8.cm]{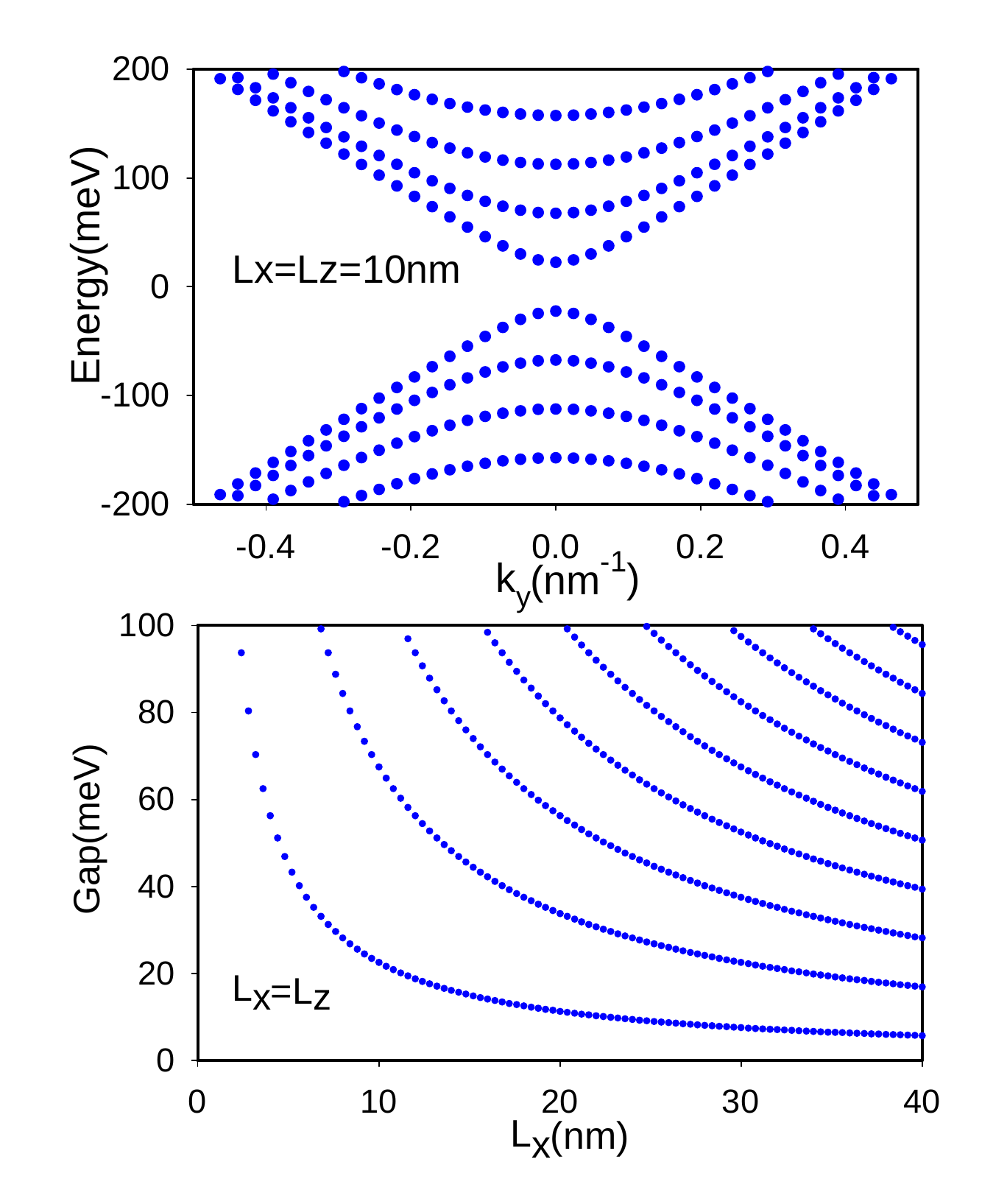}
 \caption{(Color online.) Top panel, band structure of a TI QW with $L_z$=$L_x$=$10$nm, as obtained from Eq. \ref{enky}. Bottom  panel, dependence of the
 $k_y$=0 energy levels  on the dimensions of the TI QW.}
 \label{Fig3}
\end{figure}

\section{Topological Insulator Quantum Wire}
As a first example of how these matching conditions can be used,
we analyze the electronic structure of the surface states of a  quantum wire (QW) with rectangular cross section.  For this example
we  neglect the diagonal term $E({\bf k})$ in Hamiltonian Eq. \ref{H3D}, which
breaks electron-hole symmetry.  This allows us to obtain analytical results which
are easily understood.

The  dimensions of the QW are $L_x$ and $L_z$ along the $x$-axis and $z$-axis respectively, and it is infinitely long in the $ \hat y$-direction,
so that $k_y$ is a good quantum number.
Given the momentum $k_y$ and the energy $E$,
for each surface of the QW ($\pm x,\pm z$) one may find
the corresponding electron wavefunctions
($\Psi ^{\pm x,\pm z}$), each as a linear combination of
the two solutions of the corresponding
Dirac-like surface Hamiltonians.  Thus, there are eight
coefficients that determine the
QW wavefunction, which obey four equations of the form
in Eq. \ref{Mat_Con}.
Explicitly, we define two-component wavefunctions
of the form
\begin{widetext}
\begin{eqnarray}
\varphi^{(+z)}(x)&=&
\left(
\begin{array}{c}
c^{(+z)}_{u} \nonumber \\ c^{(+z)}_{v}
\end{array} \right)_{+k_x}
A^{(+z)}e^{ik_x x}
+
\left(
\begin{array}{c}
c^{(+z)}_{u} \nonumber \\ c^{(+z)}_{v}
\end{array} \right)_{-k_x}
B^{(+z)}e^{-ik_x x} \nonumber \\
\varphi^{(+x)}(z)&=&
\left(
\begin{array}{c}
c^{(+x)}_{u} \nonumber \\ c^{(+x)}_{v}
\end{array} \right)_{+k_z}
A^{(+x)}e^{ik_z z}
+
\left(
\begin{array}{c}
c^{(+x)}_{u} \nonumber \\ c^{(+x)}_{v}
\end{array} \right)_{-k_z}
B^{(+x)}e^{-ik_z z} \nonumber \\
\varphi^{(-z)}(x)&=&
\left(
\begin{array}{c}
c^{(-z)}_{u} \nonumber \\ c^{(-z)}_{v}
\end{array} \right)_{+k_x}
A^{(-z)}e^{ik_x x}
+
\left(
\begin{array}{c}
c^{(-z)}_{u} \nonumber \\ c^{(-z)}_{v}
\end{array} \right)_{-k_x}
B^{(-z)}e^{-ik_x x} \nonumber \\
\varphi^{(-x)}(z)&=&
\left(
\begin{array}{c}
c^{(-x)}_{u} \nonumber \\ c^{(-x)}_{v}
\end{array} \right)_{+k_z}
A^{(-x)}e^{ik_z z}
+
\left(
\begin{array}{c}
c^{(-x)}_{u} \nonumber \\ c^{(-x)}_{v}
\end{array} \right)_{-k_z}
B^{(-x)}e^{-ik_z z}, \label{wf_defs}
\end{eqnarray}
\end{widetext}
where $k_x$ is the value that, when substituted into Eq. \ref{Hz}, yields
a particular energy eigenvalue $E$, $k_z$ is the analogous value for
Eq. \ref{Hx},
$(c^{(\mu)*}_{u} \, c^{(\mu)*}_{v})^{\dag}_{\pm k_{\nu}}$
are the normalized eigenvectors of these Hamiltonians,
and $A^{(\pm\mu)}$, $B^{(\pm\mu)}$ are coefficients which
must be determined by matching at the corners.
These matching conditions are
\begin{eqnarray}
\varphi ^{(+z)} (x\!=\!0) \! \! \! & =& \! \! \!
{\Large M}_{z,x} \, \varphi ^{(+x)} (z\!=\!0) \nonumber \\
\varphi ^{(-z)} (x\!=\!0) \!\!  \!& =& \! \! \!
{\Large M}_{-z,x} \,  \varphi ^{(+x)} (z\!=\!-L_z) \nonumber \\
\varphi ^{(-z)} (x\!=\!-L_x)\!\!  \! & =& \! \! \!
{\Large M}_{-z,-x} \,  \varphi ^{(-x)} (z\!=\!-L_z) \nonumber \\
\varphi ^{(+z)} (x\!=\!-L_x)\!\!  \!& =& \! \! \!
{\Large M}_{z,-x} \,  \varphi ^{(-x)} (z\!=\!0).
\label{BC_QW}
\end{eqnarray}
For a given momentum $k_y$, the matching conditions can
only all be met at particular energies $E=\epsilon _{n,k_y}$
that define the QW band structure.
If we neglect $E({\bf k})$ in Eq. \ref{H3D},
it is possible after some algebra
to find these energies analytically, with the result
\begin{equation}
\epsilon _{n,k_y} = \pm \sqrt{(A_2 k_y)^2+\left (    \pi \frac {A_1 A_2}{A_1 L_z+ A_2 L_x} (n - \frac 1 2 )\right ) ^2 } ,
\label{enky}
\end{equation}
with $n$=1,2,3..., and we have made the further simplifying assumption
that $B_1=B_2=D_1=D_2=0$.
This band structure is spin degenerate. 

Eq. \ref{enky} can be easily rationalized with a geometrical argument.
When a carrier moves in a closed loop around the quantum wire,
the matching of the wavefunctions yields the quantization condition
\begin{equation}
2(k_x L_x+k_zL_z)+\pi = 2 \pi n \, .
\label{quant}
\end{equation}
The phase $\pi$ in the left part of the quantization equation appears because of the helical nature of the carriers: when the electrons encircle the QW, the
expectation value of the Pauli matrices that appear in the   Dirac  Hamiltonians rotates by $2\pi$, so that the wavefunction acquires a Berry
phase of $\pi$. (An analogous accumulation of phases occurs in
graphene hexagonal quantum rings with discrete $120^\circ$
corners \cite{Luo_2009}.)
Moreover, the wavevectors
$k_z$ and $k_x$ are related to the energy by
\begin{equation}
E=\sqrt{(A_1 k_z)^2+(A_2 k_y)^2 }=\sqrt{(A_2 k_x)^2+(A_2 k_y)^2 } \, .
\label{enek}
\end{equation}
Combining Eqs. \ref{quant} and \ref{enek}, one obtains the band structure
Eq. \ref{enky}.

In Fig. \ref{Fig3} we plot the band structure as a function of the momentum $k_y$ for a QW with $L_z$=$L_x$=$10$nm, and the dependence of the
$k_y$=0 energy levels on the dimensions of the the QW.  Analogous
band structures for a TI QW have been obtained in a cylindrical
geometry \cite{Egger_2010}.
\begin{figure}
 \includegraphics[clip,width=8cm]{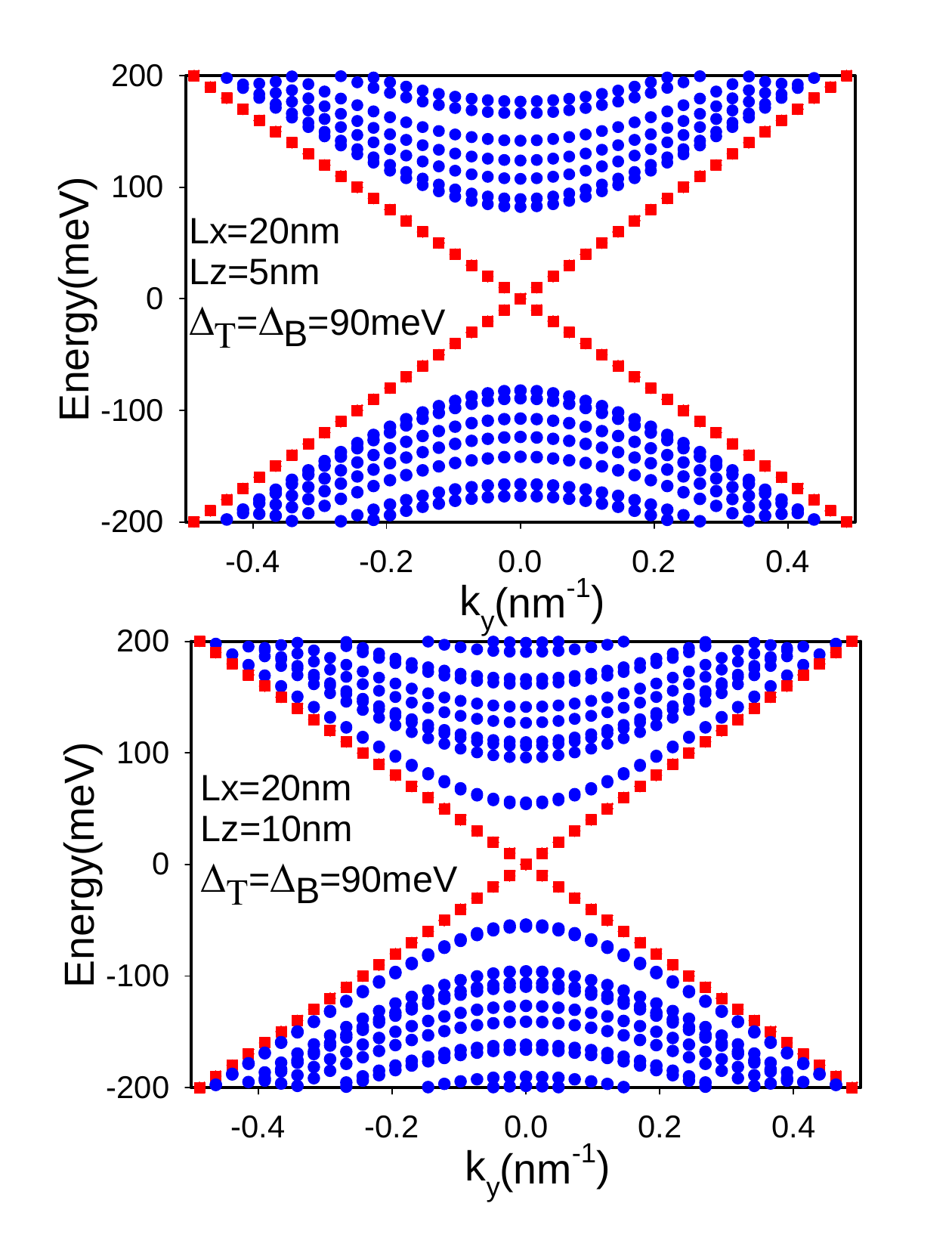}
 \caption{(Color online.) Energy bands of a TI QW in presence of  equal exchange fields $\Delta _T$=$\Delta _B$=90meV in  top and bottom surfaces.   Blue circle points are not chiral states, while red squares correspond to chiral states. Dimensions of the QW are
 $L_x$=20nm  and $L_z$=5nm (10nm), in the top (bottom) panel.}
 \label{Fig4}
\end{figure}

\begin{figure}
 \includegraphics[clip,width=8.cm]{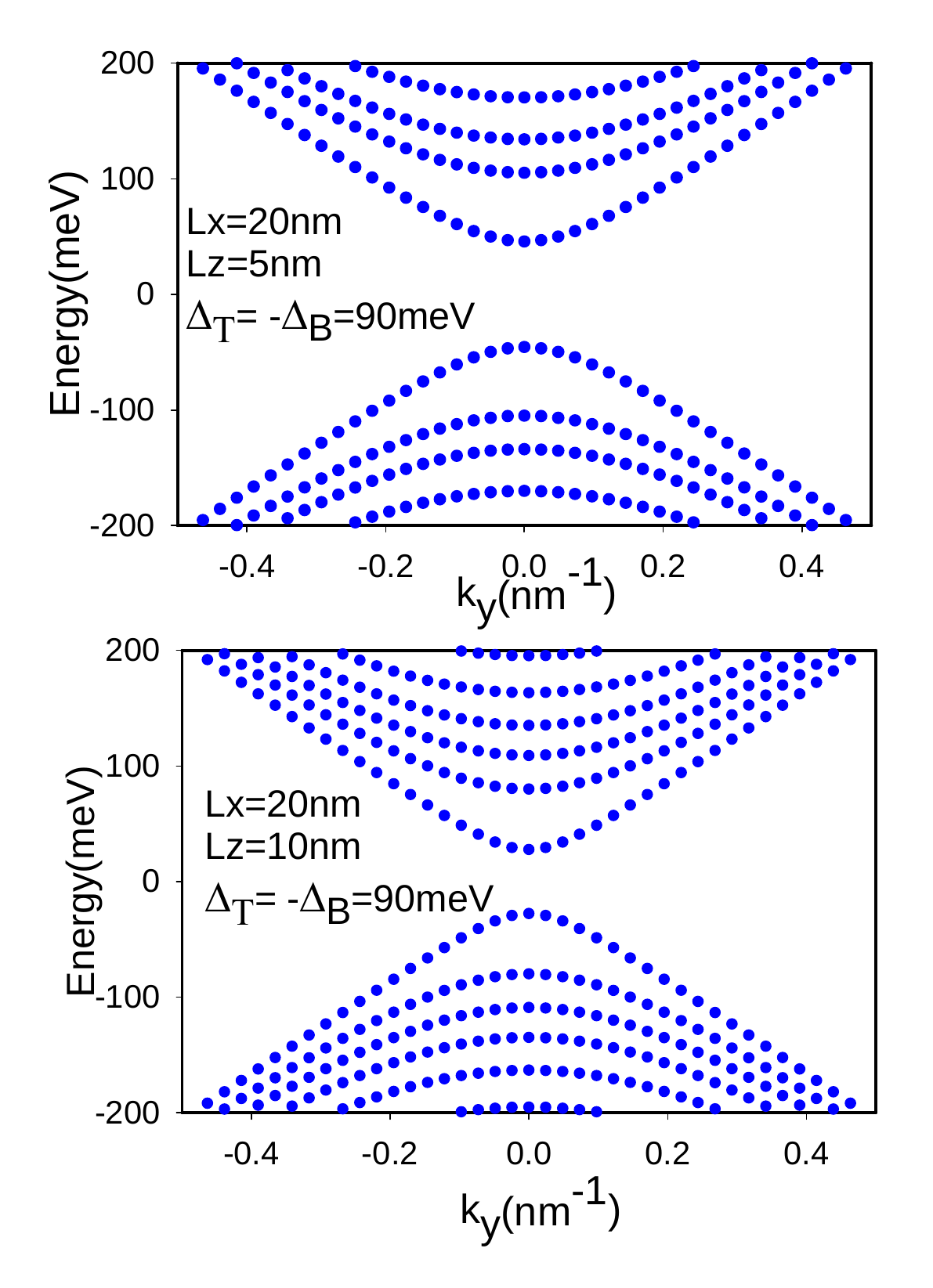}
 \caption{(Color online.) Energy bands of a TI QW in presence of  exchange fields of opposite sign in the the top and bottom surfaces, $\Delta _B$=-$\Delta_T$= 90meV.  In this configuration the total Chern number is zero and  there are no chiral states. Dimensions of the QW   are
 $L_x$=20nm  and $L_z$=5nm (10nm), in the top (bottom) panel}
 \label{Fig5}
\end{figure}

\section{Exchange fields and anomalous quantum Hall effect}
Interesting physics can be induced in these types of systems by the
introduction of time-reversal symmetry breaking perturbations on the
surfaces.  As discussed in the Introduction, this can be accomplished
by thin film ferromagnets, exchange coupled to one or more
surfaces of the system.  In particular these can couple to the
spin of the electrons without introducing orbital magnetic flux
into the Hamiltonian.

For a single such surface, for example with normal in the $\pm {\hat z}$ direction,
an exchange field $\vec\Delta $ parallel to this
opens an energy gap (see Eqs. \ref{Hz}-\ref{spinz}.)  By contrast, if
the surface has normal perpendicular to $\vec\Delta$, the spectrum
remains gapless (see Eqs. \ref{Hx}-\ref{spinx}.)
An isolated gapped surface appears to support an {\it anomalous} half integer
Hall conductivity
$\sigma _{xy}$=sign$(\Delta)\frac 1 2 \frac {e^2}h$.
As discussed in the Introduction, in real geometries for which
there must be a
top and bottom
surface, the Hall conductivity becomes integrally quantized.

In this Section we analyze a TI quantum wire of rectangular section
in presence of a $z$-polarized exchange field. $\Delta$ enters
as a mass term in the Dirac Hamiltonians for the $\pm {\hat z}$ surfaces,
but does not qualitatively modify the
Hamiltonians corresponding to the $\pm {\hat x}$ and $\pm {\hat y}$ surfaces.
Again, in order to simplify the discussion, we neglect in the Hamiltonian terms proportional to $D_1$ and $D_2$.

Fig. \ref{Fig4} illustrates the energy spectrum
of a TI QW with lateral dimensions $L_z$=20nm and $L_z$=5nm (top panel)
and $L_z$=10nm (bottom panel) in the presence of an exchange field
of magnitude 90meV. There are three kinds of states.
({i}) For energies smaller than $\sqrt{\Delta ^2+ A_2 ^2 k_y ^2}$, there are  states confined to the lateral surfaces, with energies below the gap for states on
the exchange-coupled surfaces.
The energies of these states
depend on the lateral size $L_z$ of the wire. For the values of $L_z$ and $\Delta$
illustrated in Fig.\ref{Fig4}, tunneling between states
on opposite lateral surfaces is essentially negligible,
so that these states are
nearly doubly degenerate; deviations from this are only apparent
at energies very close to $\Delta$.
({ii}) At energies larger than $\Delta$ the states extend along  the entire perimeter of the TI QW. Because time reversal symmetry is broken by the exchange field, these states are not degenerate. ({iii}) Finally, there gapless modes with linear
dispersion $\pm A_2 k_y$.  These describe chiral states moving in opposite directions on opposite lateral surfaces.

States of type (i) and (ii) are not chiral: for each state there is
a counter-propagating state on the same surface.  Impurities can induce backscattering among these states and lead to localization.
Chiral states moving in opposite directions reside on opposite surfaces,
and for a wide enough system, backscattering is negligible.
Magnetically gapped top and bottom surfaces are always connected
by surfaces with these chiral states, so that
the {\it anomalous} Hall conductivity of the system as a whole is $e^2/h$.

To gain more insight into the nature of the chiral states,
we look for a criterion that determines when they are present.
Consider a system in which the top and layers are perturbed
by exchange fields $\Delta _T$ and $\Delta _B$, respectively.
We look for wavefunctions on a single lateral surface
with momentum $k_z$=0, and energy
$E$=$sA_2 k_y$, with $s$=$\pm 1$. In this geometry the top and
bottom wavefunctions (extending into the $x-y$ plane with $x<0$)
and the lateral wavefunction (at $x=0$) have the form

\begin{eqnarray}
&\varphi^{(+z)}  =  C\left ( \begin{array} {c} 1 \\   \frac { sA_2 k_y -\Delta _T}{A_2 k_y +|\Delta _T|}  \end{array} \right )  e ^{|\Delta _T| x},\nonumber \\
&\varphi^{(-z)}  =  C ' \left ( \begin{array} {c} -1 \\   \frac { sA_2 k_y -\Delta _B}{A_2 k_y +|\Delta _B|}  \end{array} \right ) e ^{|\Delta _T| x}, \nonumber \\
&\varphi ^{(x)}  =  \alpha  \left ( \begin{array} {c} 1-s  \\ 1+s   \end{array} \right )+\beta \left ( \begin{array} {c} 1-s  \\ 1+s   \end{array} \right ).
\end{eqnarray}
Solutions with energy $\pm s A_2 k_y$ will exist if these wavefunctions satisfy  the boundary conditions Eq.\ref{Mat_Con} at the matching points ($x$=0,$z$=0) and  ($x$=0,$z$=$-L_z$),
\begin{eqnarray}
C\left ( \begin{array} {c} 1 \\   \frac { sA_2 k_y -\Delta _T}{A_2 k_y +|\Delta _T|}  \end{array} \right )
&=&\left ( \begin{array} {c} \alpha +\beta  \\ s (\alpha +\beta )  \end{array} \right ),
 \nonumber \\
C ' \left ( \begin{array} {c} -1 \\   \frac { sA_2 k_y -\Delta _B}{A_2 k_y +|\Delta _B|}  \end{array} \right )
&=&\left ( \begin{array} {c} -(\alpha +\beta) s  \\  \alpha +\beta    \end{array} \right )\,.
\label{match_chiral}
\end{eqnarray}
For top and bottom exchange fields with the same sign, the boundary conditions are only satisfied for $s$=-1. Therefore in the lateral surface (normal to $\hat x$-direction) there is a chiral state where the electrons move in the \hbox{-$\hat y$-}direction with speed $A_2$. Similar equations can be written for the opposite lateral surface, normal to the \hbox{-$\hat x$-}direction,
where the band dispersion is $A_2k_y$, and the chiral carriers also move in
the $\hat y$-direction with speed $A_2$, albeit in the opposite direction.

Finally, it is interesting to see what happens to this picture
when the exchange fields on the top and bottom surfaces
point in opposite directions, $\vec\Delta _T \cdot \vec\Delta _B <0$.
In this case Eqs. \ref{match_chiral} have no solutions, and chiral states
are not present in the system. Fig. \ref{Fig5} illustrates a full solution
of the problem as described in the last section, corroborating this structure.
This is consistent with general considerations in terms of the surface
Chern numbers:
the top and bottom surfaces have Chern number $\pm 1/2$, so that
the  system as a whole has Chern number zero. In this situation
(and in the absence of gapless lateral states) the system
does not exhibit an anomalous quantized Hall effect.

\begin{figure}
 \includegraphics[clip,width=8cm]{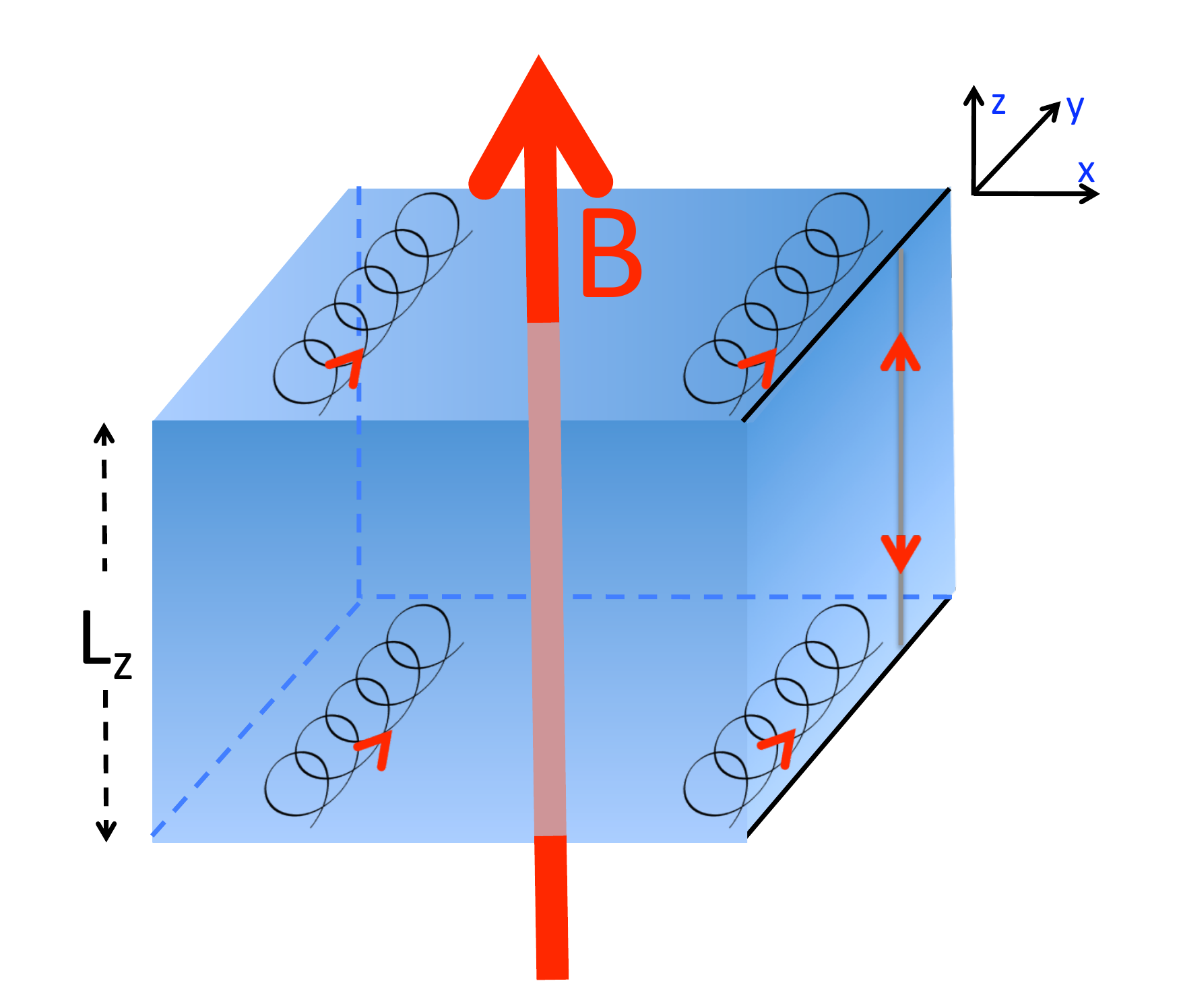} \caption{(Color online.) Schematic
 diagram of a semi-infinite thin slab of TI in presence of a perpendicular magnetic field. The slab, of thickness $L_z$, is perpendicular to the $\hat z$-direction, invariant in the $\hat y$ direction and it is defined for
 $x<0$. Carriers in the top and bottom surfaces are in the same magnetic field, whereas electrons in the lateral surface are not affected by it.
When the guiding center of  the electron motion is located away from the edge of the sample, the electronic wavefunctions of the top and bottom surfaces are those of bulk Landau levels.
When a guiding center  approaches the edge, wavefunctions on top and bottom surfaces 
become coupled through lateral surface plane waves states.}
 \label{Fig6}
\end{figure}

\begin{figure}
 \includegraphics[clip,width=8.5cm]{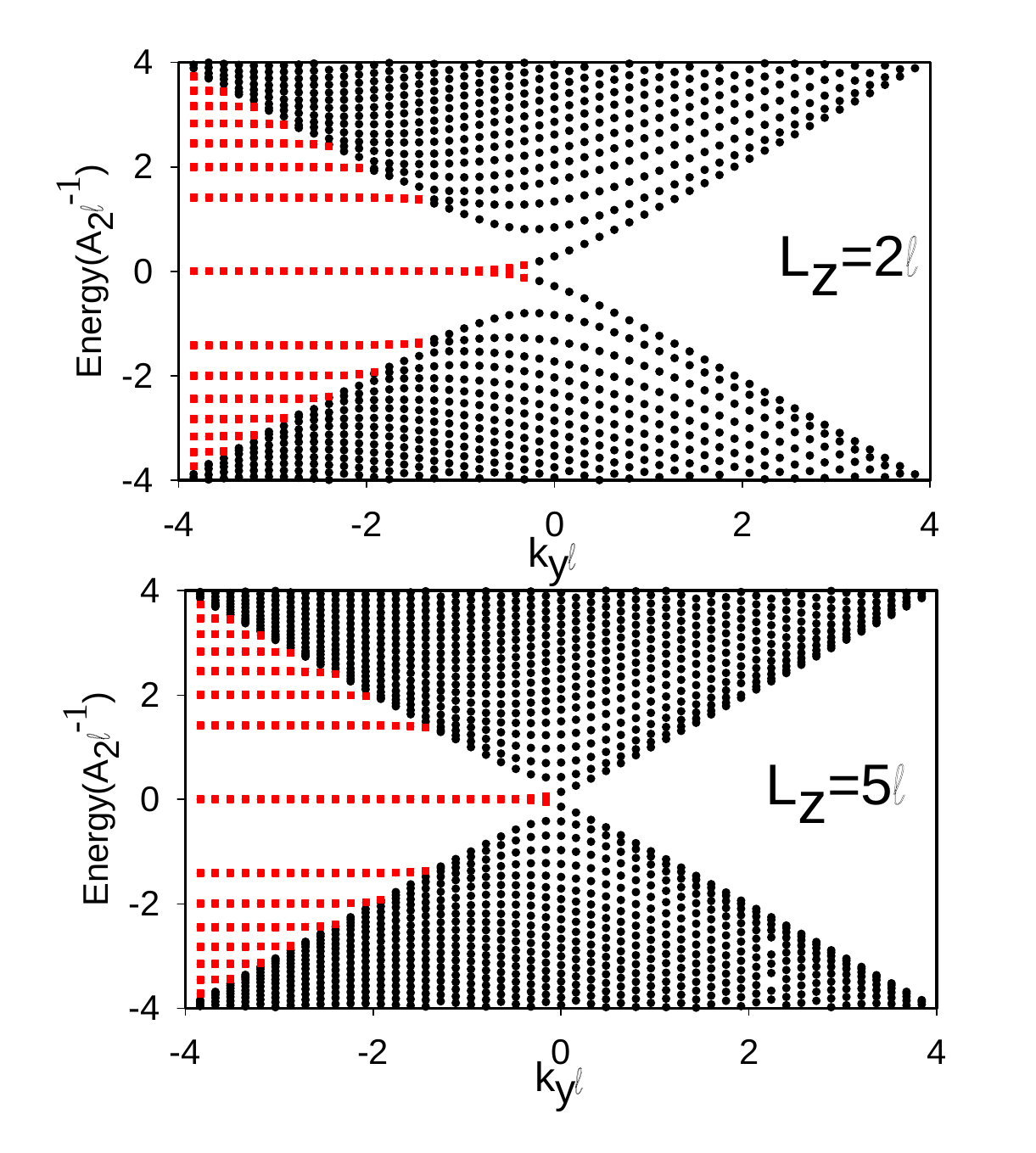}
 \caption{(Color online). Energy spectrum of a TI slab of thickness $L_z$=2$\ell$ (top) and $L_z$=5$\ell$ (bottom). Red square points correspond to wavefunctions that are mainly located in the top and bottom layers and decay exponentially in the  lateral surfaces. For $k_y \ell \ll -1$ these states evolve into
bulk Landau levels. For large values of -$k_y \ell$ the top and bottom surfaces  Landau levels are degenerate. For $k_y\ell \gtrsim -l$, top and bottom Landau levels couple through the lateral
states and the degeneracy is lifted. Black dot points correspond to states confined in the lateral surfaces. The energy spacing of these states scale as $1/L_z$\cite{Vafek_2011}.}
 \label{Fig7}
\end{figure}

\section{Landau levels, edge states and quantum Hall effect in a TI slab.}

\subsection{Energy spectrum}

In this  section we study the electronic band structure of a TI slab in the presence of a perpendicular magnetic field $B$. The magnetic field points in the $\hat z$-direction and does not affect the
motion of electrons on surfaces where this is in the plane.
We choose the Landau gauge ${\bf A}=(0,-Bx,0)$, which does not depend on the coordinate $y$, so that the wavevector $k_y$ is
a good quantum number.
In what follows we again neglect the diagonal terms involving  $E({\bf k})$ in the three dimensional Hamiltonian Eq. \ref{H3D}. Adopting $A_2 \ell ^{-1} $
as our unit of energy and $\ell=\sqrt { \hbar c/eB}$ as our unit of length,
the Hamiltonian Eq. \ref{Hz} in the presence of the magnetic field takes the form
\begin{equation}
H^{\pm z} = \pm \left(
  \begin{array}{cc}
     0 &  -\sqrt{2} \partial _z + \frac {z} {\sqrt{2}}\\
       +\sqrt{2} \partial _z + \frac {z}{\sqrt{2}} &0  \\
  \end{array}
\right),
\label{HzB}
\end{equation}
with $z=\sqrt{2} ( k_y -x)$.
The eigenvectors $(\phi_1 ^{\pm},\phi_2 ^{\pm})$
of Eq. \ref{HzB} are obtained by squaring the eigenvalue equation
$H^{\pm z} \phi = E \phi$, yielding
\begin{eqnarray}
\left ( \partial _z ^2 - \frac {z^2} 4 + \frac {E^2} 2 + \frac 1 2 \right ) \phi _1 & = &0 \\
\left (\sqrt{2} \partial z + \frac z {\sqrt{2}} \right ) \phi _1 & = & \pm \phi _2 \, .
\end{eqnarray}
Solutions of the above equations that  do not diverge at $x \rightarrow - \infty$  are
\begin{equation}
\left ( \begin{array}{c} \phi _1  ^{\pm }\\ \phi_2 ^{\pm} \end{array} \right ) = \alpha
\left ( \begin{array}{c} D _{\frac {E^2} 2 } ( \sqrt{2} (k_y -x)) \\ \pm \frac E {\sqrt{2}} D _{\frac {E^2}  2 -1 } ( \sqrt{2} (k_y -x))
\end{array} \right ),
\end{equation}
where $D_p(z)$ are parabolic cylinder functions \cite{Gradshteyn_book}.
Carriers moving on the lateral surface $\hat x$ are not affected by the magnetic field so
that the wavefunctions are eigenstates of the Hamiltonian in Eq. \ref{Hx}.
By matching of these lateral wavefunctions with those on the top and bottom surfaces
we obtain the band structure of a semi-infinite TI slab in presence of the magnetic field.
The geometry is illustrated schematically in Fig. \ref{Fig6}.

In Fig. \ref{Fig7} we plot the results of such a calculation \cite{comment_connect_to_oskar}.
For large and negative momentum $k_y$, the guiding center of the electron orbits, $k_y \ell^2$,
is located well inside the top and bottom surfaces where there is a uniform magnetic field,
and the coupling to the lateral surface
is very small.  The spectrum then consists of double degenerate Landau levels, one each
for the top and bottom layers, with energies $ \pm \sqrt{2n}  A_2 \ell ^{-1}$.
As $k_y$ increases, approaching zero from below, the Landau level wavefunctions approach
and acquire non-negligible coupling to the lateral surface states.
For the $n$th Landau level, when  $-k_y \ell \sim \sqrt{2|n|}$, this coupling becomes
important and, for $|n|>0$, the absolute value of the energy decreases.  This occurs
because the wavefunction
penetrates into the (zero-field) lateral surface, where the carriers can have a smaller kinetic
energy than in the presence of the field \cite{Masir_2008}. The $n=0$ Landau levels of the top
and bottom surfaces behave
differently because they carry no kinetic energy.  When the guiding center approaches the
junction with the lateral surface, coupling between
them becomes important and they form bonding and anti-bonding states
with the accompanying level repulsion. For $k_y\ell^2$ well inside the lateral surface,
one finds bound states due to its finite width in the $\hat{z}$ direction; the energy
spacing between these states scales as $1/L_z$ \cite{Vafek_2011}. Increasing the width
of the lateral surface (i.e., the separation between top and bottom surfaces)
generates more lateral bound states, but does not affect the separation
between Landau levels.

\begin{figure}
 \includegraphics[clip,width=8cm]{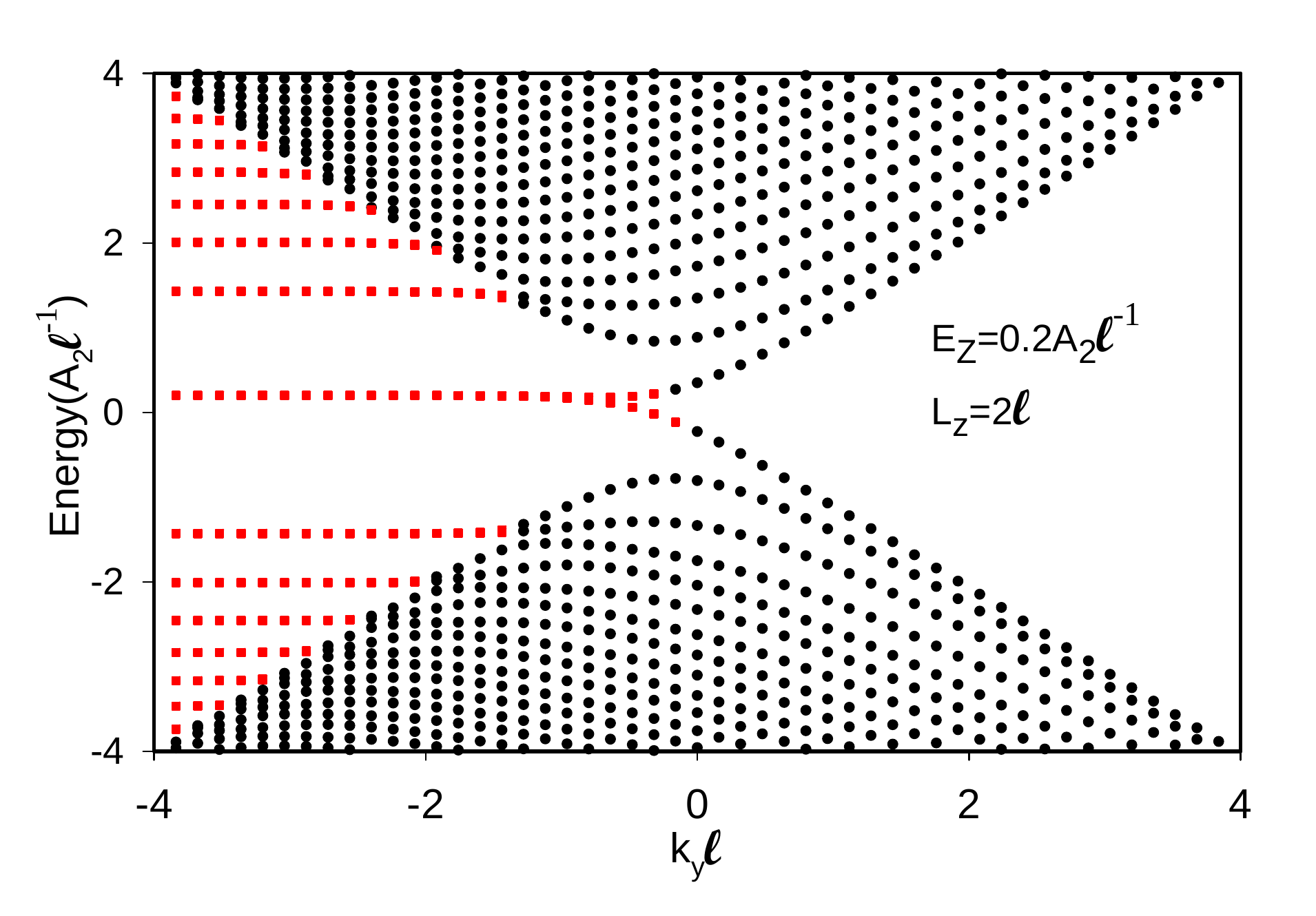}
 \caption{(Color online). Energy spectrum of a TI slab of thickness $L_z$=2$\ell$ with
 an orbital $B$-field as in \ref{Fig. 7} and a Zeeman field coupling with energy
 $E_z$=0.2$A_2 \ell ^{-1}$. }
 \label{Fig8}
\end{figure}

The discussion above neglects the coupling of the electron spin to the magnetic field.
In the $\pm {\hat z}$ surfaces there is an additional Zeeman coupling,
so the Landau level energies become
\begin{equation}
E_{\pm} = \pm \sqrt{( 2  A_2 \ell ^{-1} n  ) ^2 + E_Z ^2},
\label{LL_Dirac_Zeeman}
\end{equation}
where $E_Z= g \mu _B B /2$, with $\mu _B$ the Bohr magneton and $g$ the
effective Land\' e factor.  Note that in some topological insulators this last quantity can be
as much as fifty times larger than for free electrons \cite{Analytis_2010}.
In Fig. \ref{Fig8} we plot the band structure for a Zeeman coupling $E_z$=0.2$A_2\ell ^{-1}$.
The main effects of the Zeeman coupling are to break the electron-hole symmetry, shift the energies of the Landau levels, and to lift the degeneracy between the $n=0$ Landau levels.

\begin{figure}
 \hbox{\hspace{-1.0cm}
 \includegraphics[clip,width=10.5cm]{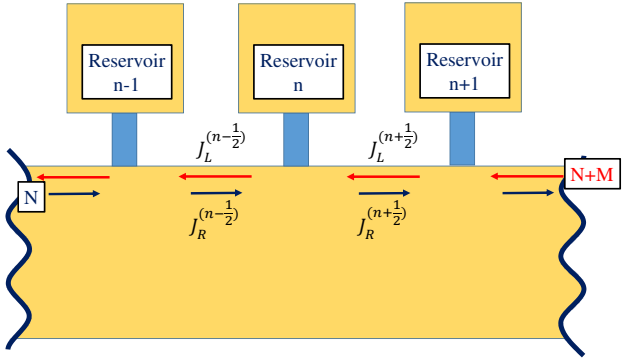}
 }
 \caption{(Color online). Model of TI slab edge connected to equilibrating leads.
 Sample edge supports $N$ right-moving channels and $N+M$ left-moving
 channels.  Left (right)-moving current between reservoirs $n-1$ and $n$
 is labeled as $J_{L(R)}^{(n-1/2)}$.
 }
 \label{edge_curr}
\end{figure}

\subsection{Quantum Hall effect}
The presence of many counterpropagating channels on the lateral surfaces can have
important consequences for the quantization of the Hall conductance in this system.
When the chemical potential is between Landau levels, for example as in the
spectra illustrated in Fig. \ref{Fig7}, it is apparent that the number of left-
and right-moving channels crossing the Fermi energy are not equal.
If there are $N+M$ channels propagating
in one direction and $N$ channels propagating in the other on each lateral surface,
when the transport on these surfaces is ballistic, the Hall conductance does
$not$ turn out to be simply $Me^2/h$ \cite{Vafek_2011}.  Moreover, the longitudinal
resistance on one of the lateral surfaces does not vanish.  In these
circumstances the system does not exhibit a quantized Hall effect.

The absence of a quantized Hall effect in these circumstances can be understood
as due to the fact that when current is injected into the system from an
ideal lead, since only channels with current directed away from the lead
can absorb this current, the distribution of currents among the channels
is out of equilibrium.  This suggests that the system will support a quantized
Hall effect if there are current-conserving mechanisms by which this
distribution can relax.  Generically this will be the case in real systems.

As a simple model, we consider a geometry as illustrated in Fig. \ref{edge_curr},
which uses phase-breaking voltage probes to equilibrate the populations
at the edges \cite{buttiker_1985,wang_2013}.
The voltage probes absorb current from each channel
with probability $\Gamma$, taken to be the same for all the channels.
Since voltage probes do not change the current down the lateral surface,
the total current absorbed by each probe is also emitted from a reservoir
at some chemical potential $\mu_n$.  The current is assumed to be injected
into each of the lateral channels with equal probability.  In this way
the voltage probes have the effect of relaxing the current into an
equilibrium distribution.  Although one can include backscattering among
channels at the edge due the leads as well as quantum interference among
the various edge and voltage probe channels in
such a model \cite{buttiker_1985}, we ignore
these possibilities to focus on phase-breaking effects.

Defining left-moving and right-moving currents between leads $n-1$ an $n$ as
$J_{n-1/2}^{(L)}$ and $J_{n-1/2}^{(R)}$, respectively, the current entering
reservoir $n$ has the form $[J_{n-1/2}^{(R)}+J_{n+1/2}^{(L)}]\Gamma$.  Current
not absorbed from a channel continues onto the next interval between reservoirs.
With the assumption that the reservoir returns all the current it absorbs back
into the lateral surface with equal probability among the outgoing channels, one finds
\begin{eqnarray}
J_{n+1/2}^{(R)}&=&(1-\Gamma)J_{n-1/2}^{(R)} +
{N \over {2N+M}}\Gamma(J_{n+1/2}^{(L)} + J_{n-1/2}^{(R)}) \nonumber \\
J_{n-1/2}^{(L)}&=&(1-\Gamma)J_{n+1/2}^{(L)} +
{{N+M} \over {2N+M}}\Gamma(J_{n+1/2}^{(L)} + J_{n-1/2}^{(R)}). \nonumber  \\
\label{current_match}
\end{eqnarray}
Note this set of equation guarantees that the net current
$J_{n+1/2}^{(R)}-J_{n+1/2}^{(L)}$ will be the same for all intervals $n$.
They can be recast into a recursion relation of the form
\begin{eqnarray}
\left(
\begin{array}{c}
J_{n+1/2}^{(R)} \\ J_{n+1/2}^{(L)}
\end{array}
\right)
= T
\left(
\begin{array}{c}
J_{n-1/2}^{(R)} \\ J_{n-1/2}^{(L)}
\end{array}
\right)
\end{eqnarray}
with
\begin{eqnarray}
T=
{1 \over {1-\gamma N}}
\left(
\begin{array}{c c}
1-\gamma(2N+M)  &  \gamma N  \\
-\gamma (N+M)   &  1
\end{array}
\right),
\end{eqnarray}
where $\gamma=\Gamma/(2N+M)$.

This recursion relation allows one to determine the currents on the top
edge, as illustrated in Fig. \ref{edge_curr},  anywhere down the length of
the sample, provided the current far to the left of the system is known.
An analogous relation can be written for the bottom edge, whose currents
depend on a boundary condition on the right.  These boundary conditions
are met at current-injecting contacts (not shown in the figure, on lateral
surfaces perpendicular to the one illustrated), and determine how the
net current down the Hall bar divides between the top and bottom lateral
surfaces.

Because the transfer matrix $T$ is independent of $n$, one may straightforwardly
determine the distribution of current in the left-moving and right-moving
channels, $\vec J_{n+1/2} = (J_{n+1/2}^{(R)} \,, J_{n+1/2}^{(L)} )^{\dag}$,
by expressing these in terms of the eigenvectors of $T$,
$\vec J_{0,t} \equiv (J_{0,t}^{(R)} \,, J_{0,t}^{(L)} )^{\dag}$,
where the corresponding eigenvalues
are $\lambda_0=1$ and $\lambda_t=1-\gamma M/(1-\gamma N)$:
$\vec{J}_{n+1/2} = a_0 \lambda_0^n \vec J_0 + a_t \lambda_t^n \vec J_t$.
The amplitudes $a_{0,t}$ are determined by the boundary conditions mentioned
above.
Since $\lambda_t<1$, the component of currents associated with $\vec J_t$ decay
away exponentially, representing a transient current distribution that
relaxes exponentially as one moves away from the current contacts.
The eigenvector $\vec J_0$ dictates the current distribution inside
the bulk.  Solving for its explicit form, one finds
$J_0^{(R)}/J_0^{(L)}=N/(N+M)$: the ratio of currents is proportional to
ratio of the number of channels.  As expected, the current relaxes into an
equilibrium distribution in which the current carried by each channel
at an edge is equal.

That the system exhibits a quantized Hall effect can easily be seen from
this last result.  For any two voltage contacts $n_{1,2}$ which are both
far from the current contacts, the transient part of
the current distribution is negligibly small \cite{wang_2013}, so that
$\vec J_{n_1+1/2} = \vec J_{n_2+1/2}= \vec J_{n_1-1/2} = \vec J_{n_2-1/2}$.
It follows that the chemical potentials
in these voltage probes must be the same, so that the measured
longitudinal resistance will vanish.  To find the Hall resistance,
we define $J^C_{T(B)}$ as the difference in current carried by each channel on the
top (bottom) edge due to currents injected/removed by the contacts far
to the right and left of the system.  On the top edge, the resulting extra current
into a voltage contact is then $\Gamma (2N+M) J^C_T$.  An equal current must
then exit from the voltage reservoir back into the system.  Assuming the reservoir
also has $2N+M$ channels, detailed balance requires the probability of
tunneling from a reservoir channel back into an edge channel is also $\Gamma$.
The extra current per channel exiting the reservoir must then
be $J^C_T$, which fixes the
change in chemical potential in the reservoir, $\delta \mu_T = h J^C_T/e^2$.
Analogous reasoning fixes the chemical potential change for voltage
probes well inside the Hall bar along the bottom edge to be
$\delta \mu_B = h J^C_B/e^2$.  Finally, recognizing that the net current
down the length of the Hall bar is $I = M(J^C_T - J^C_B) =
M{{e^2} \over h}(\delta \mu_T - \delta \mu_B)$, we arrive at a quantized
Hall conductance of $M{{e^2} \over h}$.

\section{Summary}

In this article we have studied a simplified model of surface states in topological
insulators.  The model allowed us to develop straightforward
matching conditions for states on different surfaces,
opening the possibility to understand the surface spectra of a
variety of mesoscopic systems.  Two systems were analyzed in this
formalism in detail: a quantum wire of rectangular cross-section,
and a slab geometry in the quantum Hall regime.

For the rectangular wire, one finds transverse states with a quantization
condition that reflects the helicity of the wavefunctions: an effective
two component spinor follows a circular trajectory as one moves around
a closed path around the wire, inducing a phase that prevents gapless
modes from appearing in the spectrum.  The resulting gap vanishes only
as the wire cross-sectional area becomes very large.  The application of
exchange fields on two of the surfaces changes the topological character
of the surface states by inducing a non-vanishing Chern number.  This
results in chiral states on lateral surfaces which are gapless.  In
contrast, if the exchange fields on the two surfaces are directed
antiparallel, the Chern number vanishes, and chiral states are
absent from the spectrum.

We also considered the surface spectrum of a slab in a magnetic field.
Landau level states appear on the surfaces perpendicular to the field,
which are continuously connected to zero field states on the lateral
surfaces \cite{Vafek_2011}.  The lateral states have unequal numbers
of channels propagating in opposite directions along the slab, in
direct analogy with what expects of edge states in the quantized Hall
effect.  The large number of counterpropagating edge channels
spoils the quantum Hall effect in this system in the ballistic regime.
This is due to the existence of unequilibrated populations of the
channels on a surface due to the injection of current in the system.
We found that
processes which restore the populations of the
edge modes into local equilibrium, as modeled by floating voltage
contacts along the edge,  will lead to a quantum Hall effect in the
system if voltage measurements are made sufficiently far from the current contacts.


\section{acknowledgments}
LB acknowledges fruitful discussions with Alfredo Levy Yeyati. Funding for this work was provided by MEC-Spain via grant FIS2012-33521.  HAF acknowledges support by the US-NSF through
Grant No. DMR-1005035, and by the US-Israel Binational Science Foundation through
Grant No. 2012120.

\section{\label{sec:appendix} Appendix A: Dirac Hamiltonian and metallic states in the $\hat z$ surface}
In this Appendix we outline how, starting from the three dimensional 
Hamiltonian Eq. \ref{H3D}, one
obtains the Dirac Hamiltonian describing the surface states of a TI. For concreteness
we discuss the $\hat z$-surface, but similar derivations can be carried out
for other surface orientations. In this surface the system is invariant in the
$\hat x$ and $\hat y$ direction, so that $k_x$ and $k_y$ are good quantum numbers.
States localized in the surface with energy in the bulk gap must decay exponentially
in the bulk.  Moreover, we adopt a vanishing boundary condition \cite{Silvestrov_2012}
right at the surface, $z$=$0$. We thus look for wavefunctions of the form
\begin{equation}
u(k_x,k_y,\lambda_{1,2}) \, e ^{i (k_xx+k_yy)} \left ( e ^{\lambda _1 z} - e ^{\lambda_2 z} \right ),
\label{wf_sur}
\end{equation}
where  Re $\lambda _{1,2} >0$ and $u$ is the spinor eigenstate of Hamiltonian Eq. \ref{H3D} corresponding to $k_x$, $k_y$ and $k_z = -i \lambda _{1,2}$. Note one needs to find two {\it different} values of $\lambda_{1,2}$ with the {\it same} such spinor \cite{Silvestrov_2012}, in order for all four components
of the wavefunction to vanish at $z=0$.
We are interested in the surface Hamiltonian to lowest non-trivial order in the wavevector;
therefore, in the spirit of the ${\bf k} \cdot {\bf p}$ approximation, we first obtain
the eigenstates for $k_x$=$k_y$=0, and then write the finite wavevector Hamiltonian in this basis.
Our approach differs from that of Ref. \onlinecite{Silvestrov_2012} in dropping the $k_x$ and
$k_y$ dependence of the basis states, which introduces higher order corrections in the wavevectors.
This turns out to be a considerable simplification which allows us to develop relatively simple
matching conditions, as well as to introduce a magnetic field in a straightforward way.


Substituting $i\lambda$ for $k_z$ in Eq. \ref{H3D}, the equation $\det [H_{3D}-E{\mathbb I}]=0$,
where ${\mathbb I}$ is the 4$\times$4 unit matrix, fixes
the inverse decay length $\lambda$.  Again ignoring the diagonal term in $H_{3D}$,
this yields the biquadratic equation
\begin{equation}
(E+D_1 \lambda ^2) ^2= (M_0+B_1 \lambda ^2 ) ^2 - A_1 ^2 \lambda ^2 \, .
\label{biqua}
\end{equation}
which fixes possible values of $\lambda$.
For each energy it is possible to obtain two solutions $\lambda_{1,2}$ with
Re$\lambda _{1,2} >0$. The required energy value is found by
imposing the condition $u(\lambda_{1})$=$u(\lambda_{2})$,
\begin{equation}
\left ( H^{3D} (\lambda _1)-H^{3D} (\lambda _2) \right ) u (\lambda_1) =0,
\end{equation}
which implies a further relation
\begin{equation}
(D_1^2-B_1^2)(\lambda_2 +\lambda _1) ^2 -A_1 ^2 =0\, .
\label{eq_lamb}
\end{equation}
This, together with Eq. \ref{biqua}, yields an energy eigenvalue
$E=\frac {D_1}{B_1} M_0$.  Each of the two allowed values of $\lambda$ furthermore admit
two eigenvectors of $H_{3D}$,
\begin{equation}
u  ^{+ \hat z}\! \! = \! \!\frac 1 {\sqrt{2}} \left ( \begin{array} {c}     \sqrt{ 1 \!  + \! \frac {D_1}{B_1}} \\   \! \! \! - i\sqrt{ 1\! -\! \frac {D_1}{B_1}}\\0 \\ 0 \end{array}\right )
 \, , \,
v ^{+ \hat z} \! \!=\!  \!\frac 1 {\sqrt{2}} \left ( \begin{array} {c}     0\\ 0 \\  \sqrt{ 1\! + \! \frac {D_1}{B_1}}  \\ \! \! \!   i\sqrt{ 1 \!- \!\frac {D_1}{B_1}} \end{array}\right ).
\label{basisx1}
\end{equation}
Finally, we project the three dimensional Hamiltonian Eq. \ref{H3D} in basis states of
the form in Eq. \ref{wf_sur}, using the spinors defined in Eq.
\ref{basisx1}.   To lowest  order in $k_x$ and $k_y$, this results in the Dirac Hamiltonian
\begin{equation}
H^{\hat z}  \! = \! \frac {D_1}{B_1}M_0
\! + \!   A_2 \sqrt{1 \! - \! \frac {D_1^2}{B_1^2}} \left(
  \begin{array}{cc}
     0  &  i k_x \! + \! k_y \\
      -i k_x \!+ \! k_y  &0 \\
 \end{array}
\right)
\end{equation}
as given in the text.


\begin{thebibliography}{35}
\bibitem[{com({\natexlab{a}})}]{comment_surface_states}
\bibinfo{note}{In general the surface states can depend on the in-plane
  momentum, but this will plays no role in what follows.}
\end{thebibliography}

\end{document}